# Room temperature giant magnetoresistance detection of spin hall nano-oscillator dynamics in synthetic antiferromagnetic Spin-Valve

Chunhao Li[1,2], Xiaotian Zhao[1,*], Wenlong Cai[3], Long Liu[1], Wei Liu [1,*], Zhidong Zhang[1]

[1] Shenyang National Laboratory for Materials Science, Institute of Metal Research, Chinese Academy of Sciences, Shenyang 110016, China.

[2] School of Materials Science and Engineering, University of Science and Technology of China, Shenyang 110016, China.

[3] Fert Beijing Institute, School of Integrated Circuit Science and Engineering, Beihang University, Beijing 100191, China.

## Abstract

Conventional spin Hall nano-oscillators (SHNOs) face fundamental power limitations due to the low anisotropic magnetoresistance of ferromagnetic layers. To address this, we developed a synthetic antiferromagnetic spin-valve (SAF-SV) heterostructure [Ta/NiFe/Ru/NiFe/Cu/NiFe/Hf/Pt] that enables efficient giant magnetoresistance (GMR)-based detection of SHNO dynamics at room temperature. The NiFe/Ru/NiFe layer in the spin-flop state achieves a maximum GMR ratio of 0.568% when paired with a NiFe free layer using Cu as the spacer, exhibiting complete independence of magnetic field/current orientation. Spin-torque ferromagnetic resonance (ST-FMR) verifies that the ferromagnetic resonance linewidth of the free layer can be effectively modulated by dc current through the Pt heavy metal layer, while maintaining decoupled dynamics from the SAF layer. Effective thermal management via a high-thermal-conductivity SiC substrate and an AlN capping layer has mitigated Joule heating from current shunting. Consequently, stable auto-oscillation peaks are observed at 0.82 mA, with frequencies tunable by an external magnetic field. Potential dual-mode oscillations also emerge under low fields. This work establishes a new paradigm for room-temperature, high-power spintronic oscillators, offering significant potential for neuromorphic computing and coherent RF communication applications.

* xtzhao@imr.ac.cn
* wliu@imr.ac.cn

**Introduction**

The relentless miniaturization of transistors, a cornerstone of Moore's Law, is increasingly constrained by quantum and thermal effects, posing significant challenges to the continued advancement of artificial intelligence (AI)[1,2]. In response, the development of novel computing architectures beyond the traditional Von Neumann paradigm has become imperative[3,4]. Spin Hall nano-oscillators (SHNOs) are promising candidates for next-generation computing due to their high oscillation frequencies, strong nonlinear responses, and remarkable synchronization capabilities, making them ideal building blocks for complex AI applications like vowel recognition[5,6]. Hence, Clarifying the physical mechanisms related to the operation of SHNOs and improving their performance are of great scientific and strategic importance.

In SHNOs, the charge current in the heavy metal layer is converted to a pure spin current by the spin Hall effect (SHE)[7,8]. This spin current applies a damping-like torque to the ferromagnetic (FM) layer, which drives the steady state precession of the magnetic moment[9–11]. Ultimately, the microwave electric signal is output through the anisotropic magnetoresistance (AMR) effect of the FM layer[12]. In recent years, SHNOs have made great strides in signal quality, synchronization, and miniaturization to improve their microwave output power[13–19]. However, it is still necessary to break through the lower AMR ratio of the ferromagnetic layer in order to improve its intrinsic microwave output power, which is the root of SHNOs performance. Although spin torque nanooscillators (STNOs) based on giant magnetoresistance (GMR) and tunneling magnetoresistance (TMR) have been extensively studied, their complex fabrication process and breakdown risk have limited their development[20–22]. Given that giant magnetoresistance is already capable of STNO readout, extending this readout mechanism to SHNO would be a promising strategy. Chen et al.'s pioneering study showed that a 4.8% GMR ratio provided by a structure in $Ir_{25}Mn_{75}$ (4 nm)/Co (2 nm)/Cu (4 nm)/Co (0.5 nm)/ Py(3.5 nm)/Pt (5 nm) could increase the output power of a SHNO to greater than 1 nW, which is a 200-fold improvement over AMR-based devices[23]. However, the thicker device thickness induces severe current shunting, resulting in excessive joule heating and thermal runaway,and necessitating cryogenic operation at 4.2 K. Meanwhile, both Joule heating fluctuations and stray-field interference from the reference layer significantly broaden auto-oscillation linewidths, degrading signal coherence.

Compared to the exchange-biased GMR system, spin-valve (SV) structures with synthetic antiferromagnets (SAFs) as a reference layer consisting of an additional FM free layer provide a viable alternative to GMR[24,25]. SAFs use nonmagnetic spacers to antiparallelly couple adjacent FM layers via interlayer exchange coupling (IEC)[26]. This configuration decreases stray fields through magnetic flux closure. Under applied fields, SAF spins exhibit a characteristic spin-flop transition, which is the central reason for being a reference layer[27]. SAFs exhibit robust noncollinear states over broad field ranges and enhanced thermal stability compared to exchange-biased systems, providing additional advantages for SHNOs applications. In this work, we realize room-temperature giant magnetoresistance detection of microwave signals in SHNOs with artificial antiferromagnetic spin-valve structures, providing a new option for giant magnetoresistance SHNOs. Due to the complex structure of the sample in this work, the shunt effect still exists, a silicon carbide substrate and aluminum nitride capping layer are used for thermal management, which has been proved to be an effective way to increase the microwave output power[28,29]. On this basis, we find that giant magnetoresistors with artificial antiferromagnetic spin valves have no direction dependence of current and magnetic field compared to SAFs structures, which is favorable for obtaining stable microwave signals for giant magnetoresistance detection. And the results of ferromagnetic resonance show that the ferromagnetic free layer in the spin-valve structure can be well characterized, and the precession mode of artificial antiferromagnetism has no adverse effect on it.

**Method**

All metal thin film samples were deposited on high-resistance silicon carbide substrates with (001) orientation by using DC magnetron sputtering system. The base pressure for magnetron sputtering was lower than $5.0 \times 10^{-7}$ Torr, while all layers were grown in an argon gas pressure of 3 mTorr. The deposition rate of each metal target was determined by weighing method and controlled below 0.7 Å/s. After multilayer film growth, strips and nano-bridge constriction structures were patterned using electron beam lithography and Ar-ion etching processes. A lift-off process and deposition of Ta (20 nm)/Cu (800 nm) as coplanar waveguide was used, with a bare sample size of 4 μm×5 μm. Finally, the same process was used to sputter 1 μm thick AlN on top of the device using reactive sputtering.

**Electrical characterization**

All the thin films were performed at room temperature. The ferromagnetic resonance measurements are performed using a perturbative coplanar waveguide ferromagnetic resonance system based on a lock-in amplifier. The magnetoresistance as well as the ST-FMR measurements were performed in the same self-built probe stage consisting of GSG probes from GGB Industries. The measurement system consists of a signal generator, lock-in amplifier, current source, nanovoltmeter, and electric magnet, and the DC and RF signals are separated with T-bias. PSD was measured with a high-frequency spectrum analyzer using a low-resolution bandwidth of 1 MHz.

**Result**

Figure 1(a) shows material stack configuration of all samples. The structure of sample SV is Ta(30)/NiFe(30)/Ru(8)/NiFe(30)/Cu(30)/NiFe(30)/Hf(6)/Pt(44). The samples SAF and F were used as contrast samples with the structures of Ta(30)/NiFe(30)/Ru(8)/NiFe(30)/Cu(30) and Cu(30)/NiFe(30)/Hf(6)/Pt(44). In samples SV and SAF, the bottom NiFe/Ru/NiFe forms an synthetic antiferromagnetic structure. While the addition of Hf in sample SV and sample F is to enhance the spin Hall angle and the SOT efficiency. Figure 1(b) shows the in-plane magnetization curve for all samples measured by PPMS at room temperature. For sample F, which has only a ferromagnetic free layer present, the magnetization saturated at a magnetic field of 150 mT, with a saturation magnetization strength of 0.93 T. Owing to the strong antiferromagnetic coupling between the NiFe layers in the NiFe/Ru/NiFe structure in the samples SAF and SV, the saturation magnetic field then exceeds 200 mT. For samples SAF and SV, the saturation magnetization strengths are 0.86 T and 0.74 T, respectively, which may be attributed to the separation effect of the nonmagnetic layers, which reduces the static magnetic energy between the ferromagnetic layers. Whereas, the presence of thicker non-magnetic Cu layer in sample SV leads to its lower saturation magnetization strength. When the magnetic field is lower than 200 mT, the synthetic antiferromagnetic structures in both samples SV and SAF are in the spin-flop state, and the magnetization strength continues to decrease. Subsequently, the magnetic moment of the NiFe/Ru/NiFe stack will be in the antiparallel state, when the magnetization strength remains constant until the external magnetic field approaches zero. Therefore, compared with sample SAF, the synthesized antiferromagnetic structure in sample

SV can still maintain the spin-flop state under the same magnetic field, and the ferromagnetic free layer has no significant effect on it.

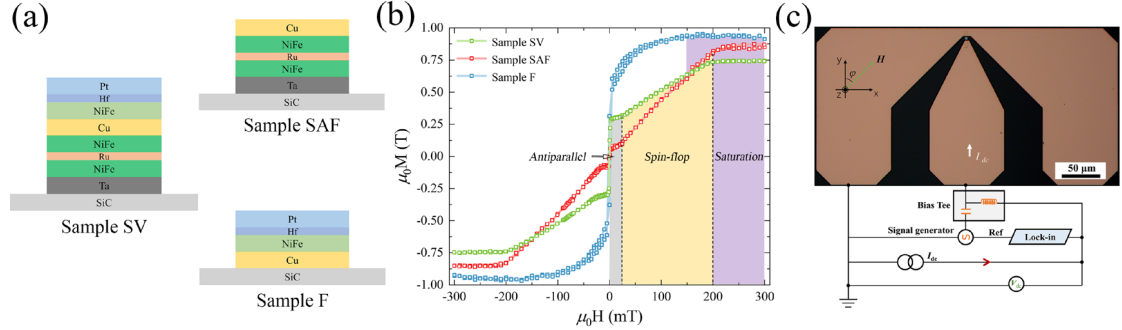

Figure.1 (a) Material stack configuration for Sample SV, Sample SAF and Sample F. (b) In-plane magnetization curve for Sample SV, Sample SAF and Sample F. (c) Schematic of magnetoresistance, FMR and ST-FMR measurement setup, the angle between the direction of the magnetic field and dc current is defined as □.

Figure 1(c) schematically shows the test setup for magnetoresistance, FMR and ST-FMR, and gives the angular relationship between the direction of the magnetic field and the direction of the dc current. All samples were tested with GSG symmetric coplanar waveguides as electrodes after patterning. To compare the GMR ratios of sample SV, we first investigated the angular dependence of the resistances of all samples, as shown in Fig. 2(a). For sample SAF and sample SV, the AMR rate of change is 0.307 % and 0.308 %, respectively, which is consistent with the results of other reference works. Due to the strong shunt effect of the Cu layer in sample F, it resulted in a small AMR rate of 0.114 %.

Since the synthetic antiferromagnetic structure also produces a giant magnetoresistance in the spin-flop state, both samples SAF and SV were measured for magnetoresistance in variable fields, shown in figure 2(b)-2(c). To better understand the variation of magnetoresistance, figure 2(d) describes the magnetic moment configurations of each ferromagnetic layer at five different magnetic field positions. For sample SAF, the magnetoresistance ratio shows opposite sign when the current is perpendicular and parallel to the magnetic field. When the current is perpendicular to the magnetic field, the magnetic moments of both ferromagnetic layers in the synthetic antiferromagnetic structure are aligned along the direction of the external magnetic field (point 1). As the magnetic field decreases, the magnetization direction of the two layers rotates in the antiparallel direction under the strong antiferromagnetic coupling of the Ru layer (point 2). During the spin-flop process, the angle between the two magnetic moments keeps getting larger, resulting in a larger giant magnetoresistance. At the same time, the angle between the magnetic moments of the two layers and the direction of the current becomes smaller, which leads to an increase in the AMR. Eventually, the magnetoresistance reaches the maximum value at the end of the spin-flip state (point 3), when the direction of the magnetic moment is parallel to the direction of the current. The synthesized antiferromagnetic structure enters the antiparallel state, which coincides with the magnetization curve of Fig. 1(b). During this period, the giant magnetoresistance remains constant. However, as the magnetic field decreases, the direction of the synthetic antiferromagnetic magnetic moment changes drastically, shifting from the current direction to the magnetic field direction, leading to a decrease in

AMR (point 4). Finally, near zero field, the magnetic moments of each layer switch in their opposite direction and the magnetoresistance undergoes an abrupt change (point 5). When the current is parallel to the direction of the magnetic field, the giant magnetoresistance changes in the same way, but the AMR changes in the exact opposite way, leading to a negative magnetoresistivity. Therefore, for sample F, the magnetoresistance is highly dependent on the angular relationship between the magnetic field and the current, which is not conducive to the conversion of magnetized auto-oscillation to electromagnetic microwave signals.

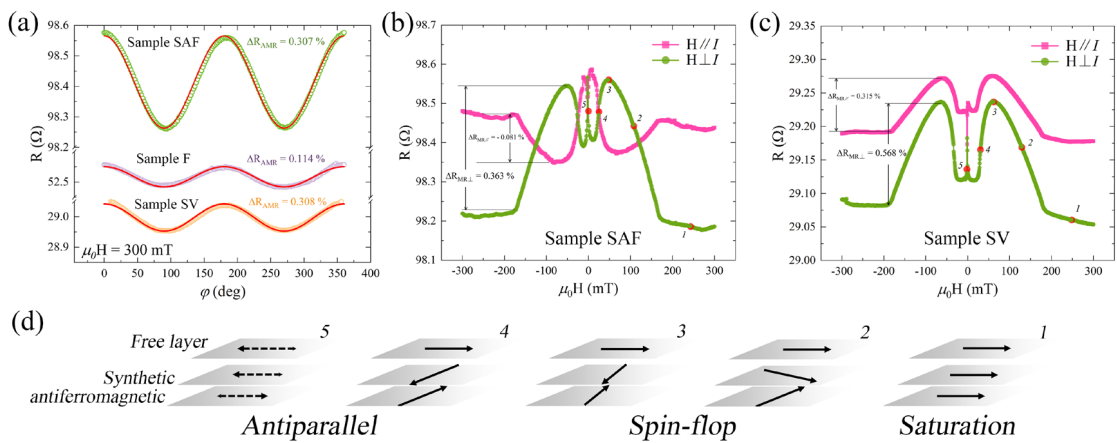

Figure.2 (a) AMR response for Sample SV, Sample SAF and Sample F as a function of the in-plane applied field angle $\varphi$. Magnetoresistance of sample SAF (b) and sample SV (c) in a magnetic field perpendicular and parallel to the current. (d) Schematic diagram of the magnetization direction of each ferromagnetic layer in synthetic antiferromagnet and ferromagnetic free layer.

For sample SV, which has structurally an additional ferromagnetic free layer on top compared to sample SAF, its magnetization configuration under the external magnetic field is shown in Fig. 2(d). When the current direction is perpendicular to the magnetic field direction, the magnetization direction change process of the synthetic antiferromagnetic structure is the same as that of the sample SAF, which results in a positive magnetoresistance ratio. Since the free layer is always along the direction of the magnetic field, the magnetization direction on both sides of the nonmagnetic Cu layer forms an angle in the spin-flop state. This further enhances the giant magnetoresistance, which is 0.568 % (0.356 % for sample SAF). The magnetoresistivity under the condition that the current direction is parallel to the magnetic field direction is changed from -0.081 % (for sample SAF) to 0.315 %. This suggests that the introduction of a free layer separated by a nonmagnetic Cu layer in the synthesized antiferromagnetic structure can significantly enhance the magnetoresistivity. The magnetoresistance is larger than the AMR at any angle of magnetic field versus current, which promises to realize the giant magnetoresistance readout of SHNO devices.

To determine the efficiency of current-driven magnetization auto-oscillation, ST-FMR measurements were performed on all samples, as shown in Figure 3. Fig. 3(a) (d) and (e) show the FMR spectra obtained from Sample SV, Sample SAF and Sample F of the bar device with an in-plane magnetic field applied from - 3000 Oe to 3000 Oe at $\varphi = 45°$, respectively. The magnetic resonance curves of the three samples were very different, especially sample SV. In order to better understand the magnetic field dependence of the resonance peaks at different microwave frequencies, the FMR spectra of sample SAF and sample F were first analyzed. As for the sample SAF, it exhibits the classical two characteristic magnetization precession modes, the acoustic mode (AC mode) with in-phase precession and the optical mode (OP mode) with out-of-phase precession. When the two magnetic layers is in the spin-flop state, the AC mode and OP mode can be described by the following Eq. 1 and Eq. 2, respectively:

$$f_{AC1} = \frac{\gamma}{2\pi}\mu_0 H_R \sqrt{1 + M_{eff}/2H_{ex}} \qquad (1)$$

$$f_{OP1} = \frac{\gamma}{2\pi}\mu_0 \sqrt{2H_{ex}M_{eff}\left[1 - \left(\frac{H_R}{2H_{ex}}\right)^2\right]} \qquad (2)$$

Here, $H_R$, $H_{ex}$, $M_{eff}$, $\gamma/2\pi$ and $\mu_0$ are the resonance field, interlayer exchange coupling field, the effective magnetization, the gyromagnetic ratio and the vaccum permeability. As the magnetic field increases (above 180 mT), the magnetic moments of the two magnetic layers align to saturate along the external magnetic field direction. Since the magnetic field needs to overcome the $H_{ex}$ and anisotropic field $H_k$, the resonance conditions change

for both resonance modes, which can be fitted using Eqs. 3 and 4:

$$f_{AC2} = \frac{\gamma}{2\pi}\mu_0\sqrt{(H_R + H_K)(H_R + H_K + M_{eff})} \quad (3)$$

$$f_{OP2} = \frac{\gamma}{2\pi}\mu_0\sqrt{(H_R + H_K - 2H_{ex})(H_R + H_K + M_{eff} - 2H_{ex})} \quad (4)$$

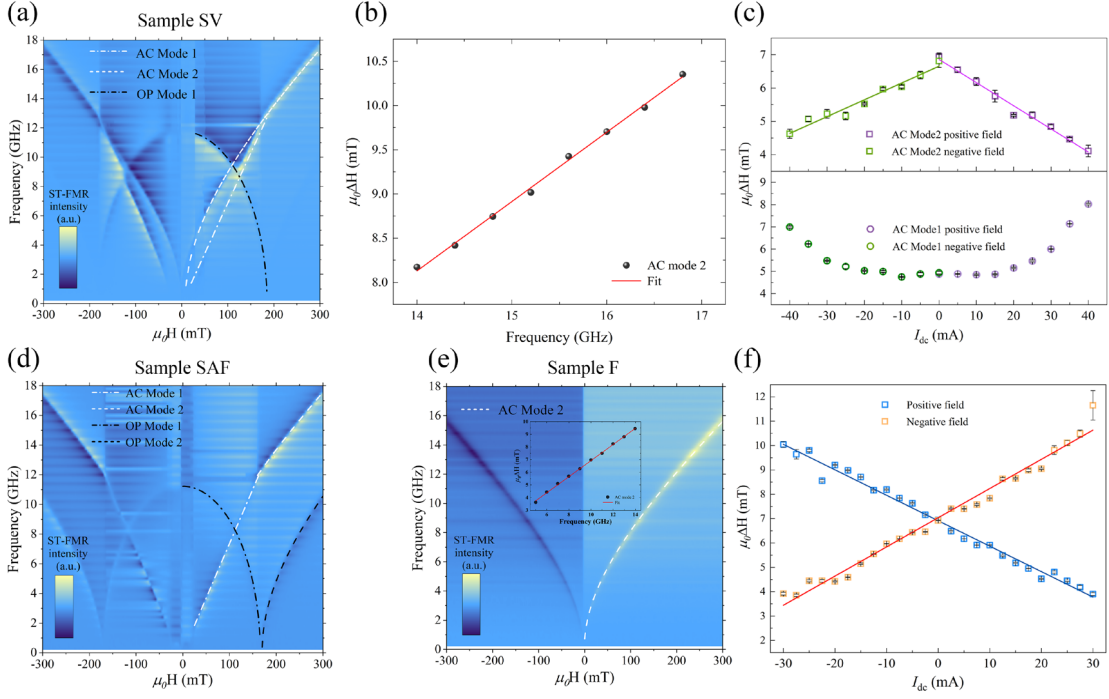

Figure.3 Sample SV (a) ST-FMR spectrum among 0.2 - 18 GHz with in-plane field angle $\varphi = 45°$. (b) Linewidth of as a function of frequency. (c) AC mode1 and AC mode2 linewidth versus dc current at 7.5GHz. Sample SAF (d) ST-FMR spectrum, $f = 0.2 - 18$ GHz, $\varphi = 45°$. Sample F (e) ST-FMR spectrum, $f = 0.2 - 18$ GHz, $\varphi = 45°$. Inset: linewidth of as a function of frequency. (f) linewidth versus dc current at 10 GHz.

According to the equation of the mode above, for sample SAF as shown in Fig. 3(d), neglecting the commonly small $H_k$ of NiFe, the fit result to be $\mu_0 M_{eff} = 0.88$ T, $\mu_0 H_{ex} = 111.3$ mT and $\gamma/2\pi = 34.14$ GHz/T. For sample F, the magnetic field dependence of the resonance can be well described by the Kittel behavior (Eq. 3), due to the presence of only a NiFe ferromagnetic free layer. Fig. 3(e) illustrates the fitting results for sample F, at parameters $\mu_0 M_{eff} = 0.72$ T, and $\gamma/2\pi = 28.30$ GHz/T. For sample SV, three curves of magnetic resonance in relation to the applied magnetic field were observed. Therefore, based on the magnetic resonance behavior of sample SAF and sample F, it is known that for sample SV, the AC2 precession modes originate from the NiFe free layer at low frequencies (below 12 GHZ), while the AC1 precession modes and OP1 precession modes originate from the artificial antiferromagnetic structure. At higher frequencies, all AC modes exhibit Kittel behavior. Accordingly, the sample SV was fitted as follows $\mu_0 M_{eff} = 0.50$ T, $\mu_0 H_{ex} = 49.7$ mT and $\gamma/2\pi = 28.00$ GHz/T. However, the OP mode2 was not observed, which may be caused by a smaller signal at the same frequency. In conclusion, both the ferromagnetic free layer and synthetic antiferromagnet isolated by nonmagnetic Cu in sample SV can individually perform magnetic resonance under microwave current, and the two layers do not affect each other.

For SHNO, it is necessary that the heavy metal layer generates spin accumulation in the neighboring ferromagnetic layer under SEH, which leads to the full compensation of the magnetic moment precession damping and the realization of the auto-oscillation. Therefore, the damping factor of the NiFe free layer and the spin Hall angle of the Pt heavy metal layer need to be considered. Since the sample SV has the NiFe free layer as an oscillator layer and its resonance behavior is well described by AC mode2, the resonance field and linewidth $\mu_0\Delta H$ can be obtained by fitting the Lorentzian function (consisting of one symmetric and one antisymmetric Lorentzian). Fig. 3(c) shows the $\mu_0\Delta H$ of the sample SV as a function of frequency. By linear fitting, $\mu_0\Delta H = \mu_0\Delta H_0 + 2\pi\alpha f / \gamma$, the nonhomogeneous broadening of $\mu_0\Delta H_{AC} = 0.18$ mT and the Gilbert damping factor of $\alpha_{AC} = 2.71\times10^{-3}$ for the AC mode are obtained. Meanwhile, the inset of Fig. 3(e) shows the extracted linewidth of sample F with a linear fit, yielding a nonhomogeneous broadening of $\mu_0\Delta H_{AC} = 0.54$ mT and the Gilbert damping of $\alpha = 3.36\times10^{-3}$. (For the samples without the introduction of the Hf layer, the Gilbert damping factor is $5.11\times10^{-3}$, presented in Supplementary Material). This indicates that the effect of the synthesized antiferromagnetic structure on the magnetic damping of the NiFe free layer is negligible. In addition, the lower damping factor may be due to the introduction of the Hf layer, which suppresses the spin-pumping (SP) effect and thus reduces the $\alpha_{SP}$ generated by spin-pumping. Then, a dc-dependent STFMR linewidth analysis was applied to modulate the damping and the SOT efficiency was estimated. For sample SV shown in Fig. 3(c), the linewidth has a linear dependence relative to the dc current, and the frequency is fixed at 7.5 GHz. When the magnetic field is along $\varphi = 105°$ ($\varphi = 285°$), the sign of the slope of the line width of the AC mode2 reflected in the NiFe free layer is negative (positive), and the positive (negative) current provides negative damping. In contrast, it does not show a significant linear relationship for AC mode1, and the trend of the linewidth with dc current is opposite to that of AC mode2. This is favorable for obtaining microwave signals at high current densities generated only by NiFe free layer progression. The effective spin Hall angle $\theta_{SH}^{eff}$ is then determined from the dc-dependent ST-FMR linewidth analysis using the following equation:

$$\theta_{SH}^{eff} = \frac{2e}{\hbar}\frac{(H + 0.5M_{eff})\mu_0 M_s t_{FM}}{\sin\varphi}\frac{\gamma}{2\pi f}\frac{\delta\Delta H}{\delta(I_{dc,HM})}A_C \quad (5)$$

where $\hbar$ and $e$ is the reduced Planck's constant and elementary charge, $M_s$ is the saturation magnetization, $t_{FM}$ is the thickness of the ferromagnetic free layer, and $A_C$ is the cross-sectional area of the measured device. The current in the HM layer, Idc,HM, is calculated using a parallel resistor model from the measured resistivity of individual FM and HM layers. The current in the HM layer, $I_{dc,HM}$, is calculated from the measured resistance of the HM layer and the total resistance using the parallel circuit model. We obtained the spin Hall angle $\theta_{SH,SV}$ = 0.110 for sample SV. As a comparison, the linear dependence of the linewidth on the dc current for sample F is also shown in Fig. 3f, which is calculated to obtain $\theta_{SH,F}$= 0.095. The effective suppression of spin pumping by the Hf layer leads to spin Hall angles that are slightly higher than those of previous studies, but still within reasonable limits. The higher spin Hall angle of sample SV may be related to the artificial antiferromagnetic configuration suppressing the dephasing during spin current diffusion. Another possible reason is that the artificial antiferromagnetic configuration plays the same spin-reflecting effect as the antiferromagnet.

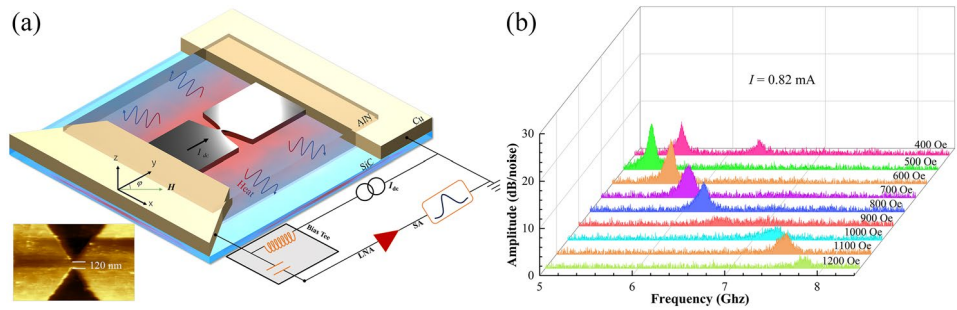

Figure.4 (a) Schematic of SHNO grown on a SiC substrate with a AlN capping layer and the connection to the measurement set-up. Inset: atomic force micrograph (AFM) of a 120 nm wide SHNO. (b) the power spectrum density as a function of magnetic field under exciting current of 0.82 mA.

With the premise of the magnetoresistance ratio and the current modulation linewidth, we next performed self-oscillation measurements of SHNO on sample SV with a nanoconstricted structure. Figure 4(a) shows schematics of the SHNO and the measurement set-up with a minimum width of 120 nm of the SHNO device, characterized by AFM. The power density spectrum in Figure 4b shows that the microwave frequency increases with increasing magnetic field, which is consistent with the Kittle equation. At the same time, the output power decreases sharply due to the increase in damping torque caused by the increase in magnetic field. The highest power density occurs at a magnetic field of 500 Oe, while two sub-oscillation peaks appear at a magnetic field of 400 Oe, which may indicate the existence of two vibration modes. Therefore, the modulation capability of the excitation current on self-oscillation needs further measurement and study.

## Conclusion

This work demonstrates the first realization of room-temperature detection of spin Hall nano-oscillator (SHNO) dynamics via the giant magnetoresistance (GMR) effect in a synthetic antiferromagnetic spin-valve heterostructure [Ta/NiFe/Ru/NiFe/Cu/NiFe/Hf/Pt]. The SAF-SV architecture achieves a GMR ratio of 0.568% while completely eliminating the orientation dependence between current and magnetic field. Spin-torque ferromagnetic resonance (ST-FMR) measurements confirm two critical breakthroughs: (1) the Hf layer suppresses spin pumping, enabling ultralow Gilbert damping ($\alpha = 2.71 \times 10^{-3}$) in the NiFe free layer, and (2) the free layer exhibits fully decoupled dynamics from the SAF reference layer with strong exchange coupling ($\mu_0 H_{ex} = 49.7$ mT). Combined with a synergistic thermal management strategy employing SiC substrates and AlN capping layers, we achieve stable auto-oscillation driven by 0.82 mA bias current in 120-nanometer nanoconstrictions. The emergence of dual-mode oscillations at 400 Oe reveals potential multimode synchronization capabilities. These breakthroughs address fundamental power scaling limitations in spintronic oscillators, establishing GMR-SHNOs as promising building blocks for non-von Neumann computing architectures. Future work will focus on investigating current-dependent power spectral density and elucidating the origin of auto-oscillation modes, alongside further optimization of material structures to achieve higher output power.